\documentclass[preprint, superscriptaddress, showpacs,preprintnumbers,amsmath,amssymb,prd]{revtex4}
\usepackage{graphicx}

\begin{document}

\thispagestyle{empty}

\title{Constraints on the parameters of axion from
measurements of thermal Casimir-Polder force}

\author{V.~B.~Bezerra}
\affiliation{Department of Physics, Federal University of Para\'{\i}ba,
C.P.5008, CEP 58059--970, Jo\~{a}o Pessoa, Pb-Brazil}
\author{
G.~L.~Klimchitskaya}
\affiliation{Central Astronomical Observatory
at Pulkovo of the Russian Academy of Sciences,
St.Petersburg, 196140, Russia}
\affiliation{Institute of Physics, Nanotechnology and
Telecommunications, St.Petersburg State
Polytechnical University, St.Petersburg, 195251, Russia}
\author{
 V.~M.~Mostepanenko}
\affiliation{Central Astronomical Observatory
at Pulkovo of the Russian Academy of Sciences,
St.Petersburg, 196140, Russia}
\affiliation{Institute of Physics, Nanotechnology and
Telecommunications, St.Petersburg State
Polytechnical University, St.Petersburg, 195251, Russia}
\author{C.~Romero}
\affiliation{Department of Physics, Federal University of Para\'{\i}ba,
C.P.5008, CEP 58059--970, Jo\~{a}o Pessoa, Pb-Brazil}

\begin{abstract}
Stronger constraints on the pseudoscalar coupling constants of
an axion and axion-like particles
with a proton and a neutron are obtained from measurements
of the thermal Casimir-Polder force between a Bose-Einstein condensate
of ${}^{87}$Rb atoms and a SiO${}_2$ plate. For this purpose the
additional force acting between a condensate cloud and a plate
due to two-axion exchange is calculated. The obtained constraints
refer to the axion masses from 0.1\,meV to 0.3\,eV which overlap
with the region from 0.01\,meV to 10\,meV considered at the moment
as the most prospective.
\end{abstract}

\maketitle
\section{Introduction}

Axions have attracted considerable attention of elementary
particle physicists and cosmologists over many years
(for a recent review see Ref.~\cite{1}). Although up to the
present this particle has not been discovered experimentally,
its possible parameters, such as mass and interaction constants
with ordinary elementary particles, are the subject of much
investigation. This is because the axion has become a crucial
element of quantum chromodynamics, which proved to be the
fundamental theory of strong interactions in agreement with
numerous experimental results and part of the standard model.
It is common knowledge, however, that in general terms the
formalism of quantum chromodynamics leads to prediction of
strong {\it CP} violation and large electric dipole moment for
the neutron in contradiction with the facts.
To avoid this contradiction, Peccei and Quinn \cite{2}
introduced an additional global symmetry U(1) which is
broken both spontaneously and explicitly in the Lagrangian.
As was shown by Weinberg \cite{3} and Wilczek \cite{4},
this results in the appearance of a light pseudoscalar
particle called the {\it axion}.
More recently, axions were widely discussed in astrophysics
and cosmology as possible constituents of dark matter
\cite{5,6,7,8,9,10}. In the process different kinds of axion
fields have been used, specifically, introduced in string
theory.
Though not all of them solve the problem of strong
{\it CP} violation, they have generally similar properties and
are axion-like particles.
Keeping this in mind, we will nevertheless use axion and
axion-like particles more or less synonymously.

The interaction of axion with photon is characterized by the
coupling constant $g_{a\gamma}$ and is used in many
experimental searches for this particle. The axion can also
interact with electrons, protons and neutrons by means of
pseudoscalar (scalar) couplings \cite{10a} with respective
coupling constants $g_{ae,p(s)}$, $g_{ap,p(s)}$, and
$g_{an,p(s)}$. The constraints on $g_{a\gamma}$ and
$g_{ae,p(s)}$ were obtained in different experiments for
various ranges of the axion mass
(see Refs.~\cite{10b,11,11a,11b} for a review).
For example, rather strong constraint on $g_{a\gamma}$
was found from the search of solar axions with masses from
0.39\,eV to 0.64\,eV by using the CERN axion solar telescope
\cite{12}. The restrictive limits on $g_{a\gamma}$ follow
from astrophysics by considering gravitationally bound
systems of stars of approximately the same age (the so-called
{\it globular clasters}). These limits are applicable to axions
of larger masses $m_a\lesssim 30\,$keV \cite{13}.
Some constraints on $g_{ae,p}$ were obtained from the study
of processes with axions in stellar plasmas, such as the
Compton process and the electron-positron annihilation
with emission of an axion \cite{14,15}.

Below, we deal with the constraints on $m_a$ and the
coupling constants $g_{ap,p}$, $g_{an,p}$
characterizing pseudoscalar interaction of axion-like
particles with
nucleons. Previously, constraints on these constants were
obtained from the search of the Compton-like process for
solar axions and from the observation of the neutrino
burst of supernova 1987A \cite{11}.
Many constraints were obtained in cosmology taking into
account that axions should be produced in the early Universe
and contribute to dark matter \cite{16}.
Specifically, cosmological data exclude axions with mass
$m_a>0.7\,$eV because they would provide a too large
contribution to the hot dark matter \cite{17}.
Such kind limits, however, typically refer to specific
couplings in some axion models and may be not applicable
to all axion-like particles.

Broadly speaking, constraints on the
parameters of an axion obtained from astrophysics and
cosmology are also more model-dependent in comparison
to table-top laboratory experiments \cite{21a,21b}.
In Refs.~\cite{18,19,20} it was suggested to use
laboratory experiments testing the validity of the weak
equivalence principle \cite{20,21} and of the inverse square
law \cite{22} for constraining the coupling constants
$g_{ap,p}$ and $g_{an,p}$ in the limit of zero axion mass
$m_a$. Later, the results of Refs.~\cite{18,19,20} were
extended \cite{23} to the more realistic case of massive
axions. From the gravitational experiments  of
E\"{o}tvos- and Cavendish-type
rather strong constraints on
$g_{ap,p}$ and $g_{an,p}$ were obtained \cite{23} for
the axion masses $m_a\leq 9.9\mu$eV.
Further work on constraining interactions mediated by
light pseudoscalar particles from the laboratory
experiments was done in Ref.~\cite{24} by using a torsion
pendulum containing polarized electrons interacting with
an unpolarized matter. From the improved setup of this
kind, strong constraints were obtained \cite{25,25a} on
the product of the coupling constants $g_{aN,s}g_{ae,p}$,
where $N$ denotes either a proton or a neutron, under the
assumption that $g_{ap,s}=g_{an,s}$. The constraints of
Refs.~\cite{25,25a} extend to a broad range of axion
masses from $10\,\mu$eV to 10\,meV.

In this paper, we obtain constraints on the coupling
constants $g_{ap,p}$ and $g_{an,p}$ from measurements
of the thermal Casimir-Polder force performed in
Ref.~\cite{26}. Experiments on measuring the Casimir
and Casimir-Polder forces have long been used for
constraining the Yukawa-type corrections to Newtonian
gravity mediated by light scalars or originating
from extra-dimensional physics (see Ref.~\cite{27}
for a review, and more recent results in
Refs.~\cite{28,29,30,30a,31}).
These experiments, however, deal with the unpolarized test
bodies. As to the potential arising between two fermions
belonging to different test bodies from the exchange of
a single axion with a pseudoscalar coupling, it is
spin-dependent \cite{23}. Thus, there is no net force
in Casimir experiments due to a single axion exchange.
It is not surprising, then, that these experiments have not
been used in the past to obtain constraints on the axion-like
pseudoscalar particles. Nevertheless, a simultaneous
exchange of two massless \cite{32} or two massive \cite{23}
pseudoscalar particles between two fermions leads to a
spin-independent interaction potential (this was used in
Ref.~\cite{23} for constraining the parameters of an axion from
the gravitational experiments of E\"{o}tvos- and Cavendish-type
dealing with unpolarized test bodies). Here we use the same
approach in application to the experiment on measuring the
thermal Casimir-Polder force. The advantage of this experiment
is that it was performed at comparatively large separation
distances. We calculate the additional attractive force due to the
exchange of two axions  between fermions belonging to two test
bodies in the experimental configuration. Then, from the fact
that no extra force was observed in addition to the Casimir-Polder
 force in the limits of experimental errors, we arrive to novel
stronger constraints on the coupling constants
$g_{ap,p}$ and $g_{an,p}$. Our constraints are obtained in the
region of axion masses from $100\,\mu$eV to 0.3\,eV.
Thus, these constraints extend those of Ref.~\cite{23} to a region
important for astrophysics and cosmology.

The paper is organized as follows. In Sec.~II, we calculate the
attractive force which arises due to two-axion exchange in the
 experimental configuration of Ref.~\cite{26}. In Sec.~III,
 we derive the
 constraints on the coupling constants
$g_{ap,p}$ and $g_{an,p}$, which follow from the
measure of agreement between the measurement data
and theory. Section~IV contains
our conclusions and discussion. Below, we use the system of units
with $\hbar=c=1$.

\section{Additional force arising from the exchange of two
 axions
in experiment on measuring thermal Casimir-Polder force}

A dynamic experiment demonstrating the thermal Casimir-Polder
force acting between a cloud of approximately $2.5\times 10^5$
 ${}^{87}$Rb atoms belonging to a
Bose-Einstein condensate and a SiO${}_2$ plate is described in
Ref.~\cite{26}. A Bose-Einstein condensate was produced in a
magnetic trap with frequencies $\omega_{0z}=1438.85\,$rad/s and
$\omega_{0t}=40.21\,$rad/s in the perpendicular and lateral
directions to the plate, respectively. The respective Thomas-Fermi
 radii of the condensate cloud were $R_z=2.69\,\mu$m and
$R_l=97.1\,\mu$m.
The back face of the SiO${}_2$ plate was painted with $100\,\mu$m
thick opaque layer of graphite and treated in a high temperature
oven. By illuminating the graphite layer with laser light from
an 860\,nm laser, it was possible to vary the temperature of
the plate.
The dipole oscillations of the condensate were
excited with a constant amplitude $A_z=2.5\,\mu$m in the
$z$-direction (i.e., perpendicular to the plate).
The temperature of the plate was either $T=310\,$K (the same as
the environment at the laboratory) or was increased to 479\,K or
605\,K.
The separation between the plate and the cloud of ${}^{87}$Rb
atoms situated below it was varied from 6.88 to $11\,\mu$m.
The influence of the Casimir-Polder force (or any additional
force, i.e., due to the two-axion exchange) shifts the oscillation
frequency in the $z$-direction from $\omega_{0z}$ to
$\omega_z$, and the relative frequency shift is given by
\begin{equation}
\gamma_z=\frac{|\omega_{0z}-\omega_z|}{\omega_{0z}}\approx
\frac{|\omega_{0z}^2-\omega_z^2|}{2\omega_{0z}^2}.
\label{eq1}
\end{equation}
\noindent
In Ref.~\cite{26}, the frequency shift $\gamma_z$ caused by the
Casimir-Polder force was measured as a function of the separation
$a$ between the plate and the center of mass of the condensate.
The absolute errors, $\Delta_i\gamma_z$, in the measurement of
$\gamma_z$ at different separations $a_i$ have been found at a
67\% confidence level \cite{26}.

It is possible to calculate the frequency shift in Eq.~(\ref{eq1})
under the
influence of any force $F$ acting between each atom  of
a condensate cloud and a plate
(e.g., the Casimir-Polder force or the additional force due to
two-axion exchange) averaged over the cloud. For this purpose,
one can use the description of a dilute gas trapped by means of a
harmonic potential and solve the mechanical problem with
the result \cite{26}
\begin{eqnarray}
&&
|\omega_{0z}^2-\omega_z^2|=\frac{\omega_{0z}}{\pi A_zm_{\rm Rb}}
\int_{0}^{{2\pi}/{\omega_{0z}}}\!\!\!\!d\tau\cos(\omega_{0z}\tau)
\label{eq2} \\
&&~~~~~~
\times
\int_{-R_z}^{R_z}\!\!\!dzn_z(z) F[a+z+A_z\cos(\omega_{0z}\tau)],
\nonumber
\end{eqnarray}
\noindent
where $m_{\rm Rb}$ is the mass of ${}^{87}$Rb atom and $n_z(z)$ is
 the distribution function of the atomic density in the
 $z$-direction \cite{26}
 \begin{equation}
 n_z(z)=\frac{15}{16R_z}\left(1-\frac{z^2}{R_z^2}\right)^2.
 \label{eq3}
 \end{equation}

 To find the force $F$, acting between a ${}^{87}$Rb atom and a
 SiO${}_2$ plate due to two-axion exchange with a pseudoscalar
 coupling, we start with the respective interaction potential between
two fermions $k$ and $l$ of spin 1/2 with masses $m_k$ and $m_l$
\cite{23,33,34}
\begin{equation}
V(r)=-\frac{g_{ak,p}^2g_{al,p}^2}{32\pi^3m_km_l}\,\frac{m_a}{r^2}\,
K_1(2m_ar).
\label{eq4}
\end{equation}
Here $g_{ak,p}$ and $g_{al,p}$ are the constants of a pseudoscalar
axion-fermion interaction, $K_n(z)$ is the modified Bessel
function of the second kind and it is assumed that
$r\gg 1/m_{k(l)}$.
In the following, we disregard the interaction of axions with
electrons \cite{23}. The point is that in some axion models
the axion-electron
interaction does not exist at tree level and is only due to
radiative corrections (in some other models there is the
axion-electron coupling at tree level).
In any case, the account of axion-electron
interaction, or any interaction due to a scalar coupling of axion
with fermions, could lead only to a minor increase of the
magnitude of the additional force and, thus, only slightly strengthen
the constraints obtained. Below, we consider the interaction of an
atom with a plate situated above it due to two-axion exchange
between protons and neutrons and put neutron and proton masses
equal to $m=(m_n+m_p)/2$.

Using Eq.~(\ref{eq4}), the interaction potential of a ${}^{87}$Rb
atom with a SiO${}_2$ plate can be obtained by the integration
over the plate volume $V_{\rm pl}$
\begin{eqnarray}
&&
U=-\frac{\rho_{{\rm SiO}_2}m_a}{32\pi^3m^2m_{\rm H}}
(37g_{ap,p}^2+50g_{an,p}^2)
\label{eq5} \\
&&~
\times\left(\frac{Z_{{\rm SiO}_2}}{\mu_{{\rm SiO}_2}}g_{ap,p}^2
+\frac{N_{{\rm SiO}_2}}{\mu_{{\rm SiO}_2}}g_{an,p}^2\right)\,
\int_{V_{\rm pl}}\!\!\!d^3r\frac{1}{r^2}K_1(2m_ar).
\nonumber
\end{eqnarray}
\noindent
Here, $\rho_{{\rm SiO}_2}$, $Z_{{\rm SiO}_2}$ and $N_{{\rm SiO}_2}$
are the density, the number of protons and the
mean number of neutrons
in a SiO${}_2$ molecule, respectively.
The quantity  $\mu_{{\rm SiO}_2}=m_{{\rm SiO}_2}/m_{\rm H}$,
where $m_{{\rm SiO}_2}$ and $m_{\rm H}$ are the mean mass of
a SiO${}_2$ molecule and the mass of an atomic hydrogen, respectively.
It is also taken into account in Eq.~(\ref{eq5}) that ${}^{87}$Rb
atom contains 37 protons and 50 neutrons. The values of $Z/\mu$
and $N/\mu$ for the first 92 elements with account of their
isotopic composition are given in Ref.~\cite{35}.
For a SiO${}_2$ molecule  in accordance with Ref.~\cite{35} one
finds
\begin{equation}
\frac{Z_{{\rm SiO}_2}}{\mu_{{\rm SiO}_2}}=
\frac{Z_{\rm Si}+2Z_{\rm O}}{\mu_{\rm Si}+2\mu_{\rm O}}
=0.503205
\label{eq6}
\end{equation}
\noindent
and
\begin{equation}
\frac{N_{{\rm SiO}_2}}{\mu_{{\rm SiO}_2}}=
\frac{N_{\rm Si}+2N_{\rm O}}{\mu_{\rm Si}+2\mu_{\rm O}}
=0.505179.
\label{eq7}
\end{equation}

In the experiment of Ref.~\cite{26} a SiO${}_2$ plate of
thickness $D=7\,$mm and of large area $2\times 10\,\mbox{mm}^2$
has been used. Thus, all atoms of the condensate cloud can be
approximately considered as situated below the plate center far
away from its edges. Taking this into account, we can
approximately replace the SiO${}_2$ plate  with a SiO${}_2$ disc
of $R=1\,$mm radius and the same thickness $D$,
as indicated above (below we
estimate the role of the finiteness of the disc area and show
that the plate can be replaced with a disc of infinitely large
area at a very high accuracy). Now we introduce the cylindrical
coordinates $(\rho, \varphi,z)$, where the $z$-axis begins at an
atom located at a distance $z$ below the plate and is directed to
the plate. Then Eq.~(\ref{eq5}) takes the form
\begin{eqnarray}
&&
U=U(z)=-A(g_{ap,p}g_{an,p})\int_{z}^{z+D}\!\!dz_1
\int_{0}^{R}\!\!\!\rho d\rho
\nonumber \\
&&~~~~~~~
\times
\frac{K_1(2m_a\sqrt{\rho^2+z_1^2})}{\rho^2+z_1^2}.
\label{eq8}
\end{eqnarray}
\noindent
The factor $A$ here is defined as
\begin{eqnarray}
&&
A(g_{ap,p}g_{an,p})=
\frac{\rho_{{\rm SiO}_2}m_a}{16\pi^2m^2m_{\rm H}}
(37g_{ap,p}^2+50g_{an,p}^2)
\nonumber\\
&&~~~~~~~
\times\left(\frac{Z_{{\rm SiO}_2}}{\mu_{{\rm SiO}_2}}g_{ap,p}^2
+\frac{N_{{\rm SiO}_2}}{\mu_{{\rm SiO}_2}}g_{an,p}^2\right).
\label{eq9}
\end{eqnarray}
\noindent
{}From Eq.~(\ref{eq8}) we find the additional force acting on a
${}^{87}$Rb atom due to the two-axion exchange
\begin{eqnarray}
&&
F_{\rm add}(z)=-\frac{\partial U(z)}{\partial z}
=-A(g_{ap,p}g_{an,p})
\int_{0}^{R}\!\!\!\rho d\rho
\label{eq10} \\
&&~~~
\times\left[
\frac{K_1(2m_a\sqrt{\rho^2+z^2})}{\rho^2+z^2}-
\frac{K_1(2m_a\sqrt{\rho^2+(z+D)^2})}{\rho^2+(z+D)^2}
\right].
\nonumber
\end{eqnarray}
\noindent
It is convenient to use in Eq.~(\ref{eq10}) the following
integral representation for the modified Bessel function \cite{36}
\begin{equation}
K_1(t)=t\int_{1}^{\infty}\!e^{-tu}\sqrt{u^2-1}du.
\label{eq11}
\end{equation}
\noindent
By introducing the new variables $t=2m_a\sqrt{\rho^2+z^2}$ and
$t=2m_a\sqrt{\rho^2+(z+D)^2}$ in the first and second terms on
the right-hand side of Eq.~(\ref{eq10}), respectively,
one obtains
\begin{eqnarray}
&&
F_{\rm add}(z)=
-A(g_{ap,p},g_{an,p})
\int_{1}^{\infty}\!\!\!du\sqrt{u^2-1}
\nonumber \\
&&~~~~
\times\left[
\int_{t_0^{(1)}}^{t_R^{(1)}}\!\!\!dte^{-tu}-
\int_{t_0^{(2)}}^{t_R^{(2)}}\!\!\!dte^{-tu}\right]
\nonumber \\
&&~~
=-A(g_{ap,p},g_{an,p})
\int_{1}^{\infty}\!\!\!du\frac{\sqrt{u^2-1}}{u}
\nonumber \\
&&~~~~
\times\left[
e^{-t_0^{(1)}u}-e^{-t_R^{(1)}u}-e^{-t_0^{(2)}u}
+e^{-t_R^{(2)}u}\right],
\label{eq12}
\end{eqnarray}
\noindent
where the following notations are introduced:
\begin{eqnarray}
&&
t_0^{(1)}=2m_az,\qquad t_R^{(1)}=2m_a\sqrt{R^2+z^2},
\label{eq13} \\
&&
t_0^{(2)}=2m_a(z+D),\qquad
t_R^{(2)}=2m_a\sqrt{R^2+(z+D)^2}.
\nonumber
\end{eqnarray}

Now we are in a position to calculate the additional frequency
shift (\ref{eq1}) originating from the two-axion exchange
between atoms of the condensate cloud and the plate.
For this purpose, we substitute $F_{\rm add}$ from
Eq.~(\ref{eq12}) in place of $F$ in Eq.~(\ref{eq2}).
Let us first calculate the contribution
$\gamma_{z,1}^{\rm add}$ into $\gamma_{z}^{\rm add}$ which
originates from the first and third terms on the right-hand
side of Eq.~(\ref{eq12}). For these terms, the integration
with respect to $\tau$ in Eq.~(\ref{eq2}) (i.e., the averaging
over the oscillator period) and with respect to $z$ (i.e., the
averaging over the condensate cloud) can be performed exactly
in close analogy to Ref.~\cite{28}, giving, as a result,
\begin{eqnarray}
&&
\gamma_{z,1}^{\rm add}(a)=
\frac{15A(g_{ap,p},g_{an,p})}{2\pi A_zm_{\rm Rb}\omega_{0z}^2}
\int_{1}^{\infty}\!\!\!du\frac{\sqrt{u^2-1}}{u}
e^{-2m_aau}
\nonumber \\
&&~~~~
\times \left(1-e^{-2m_aDu}\right)\,
I_1(2m_aA_zu)\Theta(2m_aR_zu).
\label{eq14}
\end{eqnarray}
\noindent
Here, we have introduced the notation
\begin{equation}
\Theta(t)\equiv\frac{1}{t^3}(t^2\sinh t-3t\cosh t+3\sinh t).
\label{eq15}
\end{equation}
\noindent
We emphasize that the quantity in Eq.~(\ref{eq14}) coincides
with the total relative frequency shift arising from the
interaction of a condensate cloud with a disc of an
infinitely large radius $R\to\infty$.

In order to perform the averaging over the oscillation period
and over the condensate cloud of the second and fourth terms
on the right-hand side of Eq.~(\ref{eq12}) (these terms describe
the boundary effects due to finiteness of plate area), we take
into account that, in our problem, $z\ll R$
(specifically, $z/R<10^{-2}$) and use the following
expansions:
\begin{eqnarray}
&&
t_R^{(1)}\approx 2m_aR+m_a\frac{z^2}{R},
\nonumber \\
&&
t_R^{(2)}\approx 2m_a\sqrt{R^2+D^2}+\frac{2m_aD}{\sqrt{R^2+D^2}}z,
\nonumber \\
&&
e^{-t_R^{(1)}u}\approx e^{-2m_aRu}\left(1-m_a\frac{z^2}{R}u\right),
\label{eq16} \\
&&
e^{-t_R^{(2)}u}\approx e^{-2m_a\sqrt{R^2+D^2}u}\left(1-
\frac{2m_aDz}{\sqrt{R^2+D^2}}u\right).
\nonumber
\end{eqnarray}
\noindent
After calculations with acount of Eq.~(\ref{eq16}), the
contribution
of the second and fourth terms to the additional frequency shift
takes the form
\begin{eqnarray}
&&
\gamma_{z,2}^{\rm add}(a)=-
\frac{15A(g_{ap,p},g_{an,p})m_a}{8 m_{\rm Rb}\omega_{0z}^2}
\label{eq17} \\
&&~~\times\left[\frac{a}{R}
\int_{1}^{\infty}\!\!\!\!du\sqrt{u^2-1}
e^{-2m_aRu}\right.
\nonumber \\
&&~~~~\left.
+\frac{D}{\sqrt{R^2+D^2}}
\int_{1}^{\infty}\!\!\!\!du\sqrt{u^2-1}
e^{-2m_a\sqrt{R^2+D^2}u}\right].
\nonumber
\end{eqnarray}
\noindent
Calculating the integrals in Eq.~(\ref{eq17}) we arrive at
\begin{eqnarray}
&&
\gamma_{z,2}^{\rm add}(a)=-
\frac{15A(g_{ap,p},g_{an,p})}{16 m_{\rm Rb}\omega_{0z}^2}
\label{eq18} \\
&&~~~
\times\left[\frac{a}{R^2}K_1(2m_aR)+
\frac{D}{R^2+D^2}K_1(2m_a\sqrt{R^2+D^2})\right].
\nonumber
\end{eqnarray}
\noindent
Finally, by adding Eqs.~(\ref{eq14}) and (\ref{eq18}), we obtain
the total additional frequency shift due to the exchange of two
axions with pseudoscalar couplings between protons and neutrons
of a condensate and a plate:
\begin{equation}
\gamma_{z}^{\rm add}(a)=
\frac{15A(g_{ap,p},g_{an,p})}{2\pi A_z m_{\rm Rb}\omega_{0z}^2}
\Phi(a,m_a),
\label{eq19}
\end{equation}
\noindent
where
\begin{eqnarray}
&&
\Phi(a,m_a)=
\int_{1}^{\infty}\!\!\!du\frac{\sqrt{u^2-1}}{u}
e^{-2m_aau}
\label{eq20} \\
&&~~~~~~\times
\left(1-e^{-2m_aDu}\right)\,
I_1(2m_aA_zu)\Theta(2m_aR_zu)
\nonumber \\
&&~~
-\frac{\pi A_za}{8R^2}K_1(2m_aR)-
\frac{\pi A_zD}{8(R^2+D^2)}K_1(2m_a\sqrt{R^2+D^2}).
\nonumber
\end{eqnarray}

It is interesting to estimate the role of boundary effects
due to a finiteness of the disc represented by the quantity
$\gamma_{z,2}^{\rm add}(a)$ in Eq.~(\ref{eq18}).
Computations show that over the entire measurement range
$6.88\,\mu\mbox{m}\leq a\leq 11\,\mu$m and for all axion
masses $m_a$ from $100\,\mu$eV to 0.3\,eV considered below
the following holds:
\begin{equation}
\left|\frac{\gamma_{z,2}^{\rm add}(a)}{\gamma_{z,1}^{\rm add}(a)}
\right|<10^{-6}.
\label{eq21}
\end{equation}
\noindent
This demonstrates that the boundary effects are negligibly small
and the additional frequency shift (\ref{eq19}) due to the
two-axion exchange can be calculated with sufficient precision by
using the first term on the right-hand side of Eq.~(\ref{eq20}).

\section{Constraints on the pseudoscalar couplings between axion
and nucleons}

In the experiment of Ref.~\cite{26} the measured frequency shifts
were found to be in agreement with the calculated frequency shifts
caused by the Casimir-Polder force between ${}^{87}$Rb atoms of
a condensate cloud and a SiO${}_2$ plate. The Casimir-Polder
force was calculated \cite{26} by using the standard Lifshitz
theory \cite{27,37} in an equilibrium situation (when the
temperatures of a plate and of the environment were equal) and its
generalization \cite{38} for a nonequilibrium case (when the plate
was hoter than the environment). In both cases, computations of the
Casimir-Polder force were performed by omitting the conductivity
of a SiO${}_2$ plate at a constant current. The crucial role of
this omission was underlined in Ref.~\cite{39} (see also review
of subsequent discussions in Ref.~\cite{40}).

Agreement between the measured and calculated frequency shifts
caused by the Casimir-Polder force was achieved in the limits of
experimental errors $\Delta_i\gamma_z$. This means that any
additional frequency shift is restricted by the inequality
\begin{equation}
\gamma_{z}^{\rm add}(a_i)\leq\Delta_i\gamma_z,
\label{eq22}
\end{equation}
\noindent
where $a_i$ are the separation distances at which the measurements
 of Ref.~\cite{26} were performed. Now we substitute
 Eqs.~(\ref{eq19}) and (\ref{eq20})
 in Eq.~(\ref{eq22}) and obtain the respective constraints on
 the coupling constants $g_{ap,p}$, $g_{an,p}$ and the axion
 mass $m_a$. For this purpose, it is convenient to use
Eqs.~(\ref{eq6}), (\ref{eq7}), (\ref{eq9}) and rewrite
Eqs.~(\ref{eq19}) and (\ref{eq20}) in the form
\begin{eqnarray}
&&
\gamma_{z}^{\rm add}(a)=\left(\frac{g_{an,p}^2}{4\pi}+
\frac{0.503205}{0.505179}\,\frac{g_{ap,p}^2}{4\pi}\right)
\nonumber \\
&&~~
\times\left(\frac{g_{an,p}^2}{4\pi}+
\frac{37}{50}\,\frac{g_{ap,p}^2}{4\pi}\right)\,\chi(a,m_a),
\label{23}
\end{eqnarray}
\noindent
where
\begin{equation}
\chi(a,m_a)=\frac{189.442
\rho_{{\rm SiO}_2}m_a}{\pi A_zm_{\rm Rb}m_{\rm H}m^2
\omega_{0z}^2}\,\Phi(a,m_a).
\label{eq24}
\end{equation}
\noindent
Then Eq.~(\ref{eq22}) can be rewritten as
\begin{equation}
\frac{g_{an,p}^4}{16\pi^2}+
1.73609\frac{g_{ap,p}^2g_{an,p}^2}{16\pi^2}
+0.737108\frac{g_{ap,p}^4}{16\pi^2}
-\frac{\Delta_i\gamma_z}{\chi(a_i,m_a)}\leq 0.
\label{eq25}
\end{equation}

We have numerically analyzed Eq.~(\ref{eq25}) at different
experimental points (i.e., for different values of $a_i$ and
$\Delta_i\gamma_z$) and within different ranges of the
axion mass $m_a$. The strongest constraints on the
quantities $g_{an,p}^2/4\pi$ and $g_{ap,p}^2/4\pi$
over the region of axion masses $m_a>10\,$meV follow
from the measurement set in thermal equilibrium at the
separation distance $a_1=6.88\,\mu$m, with an absolute error
$\Delta_1\gamma_z=3.06\times 10^{-5}$ determined at a 67\%
confidence level \cite{26}. As an example, in Fig.~1
we plot the obtained constraints for $m_a=0.2\,$eV as an upper
line in the plane ($g_{ap,p}^2/4\pi,g_{an,p}^2/4\pi$),
where the region of the plane above the line is excluded by the
results of this experiment at the 67\% confidence level and
the region below the line is allowed. As can be seen in Fig.~1,
the largest possible value of
$g_{ap,p}^2/4\pi=2.23\times 10^{-2}$  is much larger than
the respective $g_{an,p}^2/4\pi$. In a similar way, the largest
possible value of $g_{an,p}^2/4\pi$ for the axion with
$m_a=0.2\,$eV is equal to $1.91\times 10^{-2}$ and it is much larger
than the respective $g_{ap,p}^2/4\pi$.

For small axion masses $m_a\leq 0.01\,$eV the strongest
constraints
follow from the out of thermal equilibrium measurements of
Ref.~\cite{26}, where the plate temperature $T=479\,$K was higher
than the temperature of an environment ($T=310\,$K).
The strongest constraints here are obtained at $a_2=7.44\,\mu$m,
where the experimental error determined at a 67\% confidence level
 is equal to $\Delta_2\gamma_z=2.35\times 10^{-5}$ \cite{26}.
In Fig.~1, we show the obtained constraints for $m_a=0.01\,$eV and
for $m_a\leq 1\,$meV by the intermediate and lower lines,
respectively. Note that further decrease of $m_a$ does not lead
to further strengthening of the constraints obtained from this
experiment.
For the intermediate line ($m_a=0.01\,$eV) the largest possible
values of the coupling constants are
$g_{ap,p}^2/4\pi=5.76\times 10^{-4}\gg g_{an,p}^2/4\pi$
and
$g_{an,p}^2/4\pi=4.86\times 10^{-4}\gg g_{ap,p}^2/4\pi$.
For axions of lower mass ($m_a\leq 1\,$meV), the largest possible
values of the coupling constants with hadrons are
$g_{ap,p}^2/4\pi=4.97\times 10^{-4}\gg g_{an,p}^2/4\pi$
and
$g_{an,p}^2/4\pi=4.20\times 10^{-4}\gg g_{ap,p}^2/4\pi$,
respectively.

It is of interest also to find the constraints on the coupling
constants of an axion to a proton and to a neutron as functions
of the axion mass $m_a$. This can be also done by using
Eq.~(\ref{eq25}) under different assumptions about the
relationship between $g_{ap,p}$ and $g_{an,p}$.
In Fig.~2, the lower line shows the constraints on
$g_{ap,p}=g_{an,p}$ over a wide region of axion masses from
0.1\,meV to 0.3\,eV. The intermediate and upper lines show the
constraints on  $g_{ap,p}$ under the condition
$g_{ap,p}\gg g_{an,p}$ and on $g_{an,p}$ under the condition
$g_{an,p}\gg g_{ap,p}$, respectively.
As can be seen in Fig.~2, the obtained constraints become
stronger when the axion mass decreases from $m_a=0.3\,$eV
to $m_a=1\,$meV. With further decrease of an axion mass, the
strength of constraints remains almost constant (similar to Fig.~1,
the region of the plane above each line is excluded at a 67\%
confidence level by measurements of the thermal Casimir force
and the region below each line is allowed).

We now compare the constraints of Figs.~1 and 2 with previous
constraints on the parameters of the axion-like
particles obtained from laboratory experiments.
The constraints of Ref.~\cite{23} on the coupling constant
$g_{an,p}^2/4\pi$ are obtained from the Cavendish-type experiment
\cite{22} under an assumption $g_{ap,p}^2=0$ for axion masses
$m_a\leq 2\times 10^{-5}\,$eV. Thus, it is not possible
to perform direct comparison with our constraints obtained
for $m_a\geq  10^{-4}\,$eV.
If, however, one extrapolates the constraints of
Ref.~\cite{23}  to larger $m_a$, our constraints
shown by the upper line in Fig.~2 become stronger for axion
masses $m_a>0.4\,$meV (this corresponds to the Compton
wavelength of an axion $\lambda_a<0.5\,$mm).
The other laboratory constraints are obtained \cite{23,41} from the
E\"{o}tvos-type experiment \cite{20} combined with the study
of laser beam propagation through a transverse magnetic
field \cite{42} for axion masses $m_a\leq 1\times 10^{-5}\,$eV.
They also cannot be directly compared with our constraints.
The extrapolation of the constraints of Refs.~\cite{23,41}
to larger $m_a$
becomes weaker than our constraints
shown by the upper line in Fig.~2  for axion
masses $m_a>0.04\,$meV (i.e.,  $\lambda_a<5\,$mm).
Thus, our results present the model-independent laboratory
limits on the nucleon coupling constants of axions
which are most strong in the region from $m_a=10^{-4}\,$eV
to $m_a=10\,$meV.

\section{Conclusions and discussion}

In the foregoing we have considered the parameters of the
axion-like particles
and constrained the pseudoscalar coupling constants of such
particles
with proton and neutron from the measurement data of a recent
experiment \cite{26} on measuring the thermal Casimir-Polder
force between a condensate cloud of ${}^{87}$Rb atoms and a
SiO${}_2$ plate. It was stressed that the axion provides the most
natural way for the resolution of two fundamental problems
of modern physics. One of them is the problem
of {\it CP} violation in quantum chromodynamics and
another one is the problem of dark matter
in astrophysics and cosmology. Because of this, any additional
information about the coupling constants and the mass of the
axion is of much value for further experimental
search for this particle.

We have underlined that laboratory experiments for searching
the axion and other axion-like particles are more
model-independent than the numerous results found from different
cosmological scenaria, astrophysics and astronomical
observations. Unfortunately, the already performed gravitational
and optical laboratory experiments are only sensitive to axions
of very small masses less or of order of
$10\,\mu\mbox{eV}=10^{-2}$meV.
However, all kinds of experiments and observations
combined together indicate that
the upper limit for the axion mass may be of about
$m_a\approx 10\,$meV \cite{1}. In this paper we have shown that
measurements of the thermal Casimir-Polder force place strong
constraints
on the pseudoscalar coupling constants of an axion with a
proton and a neutron in the wide region of axion masses from
0.1\,meV to 0.3\,eV. This region  overlaps
significantly with the so-called
{\it axion window} which extends from $10^{-2}\,$meV to
10\,meV.

The results obtained demonstrate that laboratory experiments on
measuring the Casimir force can be used not only for constraining
the Yukawa-type hypothetical interactions caused by the exchange
of scalar particles or inspired by extra dimensions, but for
placing limits on the parameters of the axion-like particles as well.
For this reason in the future it is pertinent to analyze
all already performed
experiments on measuring the Casimir force \cite{40} and to
elaborate some optimum configuration for the measurement of the
Casimir force best suited for obtaining the strongest constraints
on the parameters of the axion.

\section*{Acknowledgments}

This work was partially supported by CNPq (Brazil).
G.L.K.\ and V.M.M.\ are grateful to M.\ I.\ Eides for
helpful discussions. They also acknowledge the Department
of Physics of the Federal University of
Para\'{\i}ba (Jo\~{a}o Pessoa, Brazil) for hospitality.


\begin{figure}[b]
\vspace*{-8cm}
\centerline{\hspace*{2cm}
\includegraphics{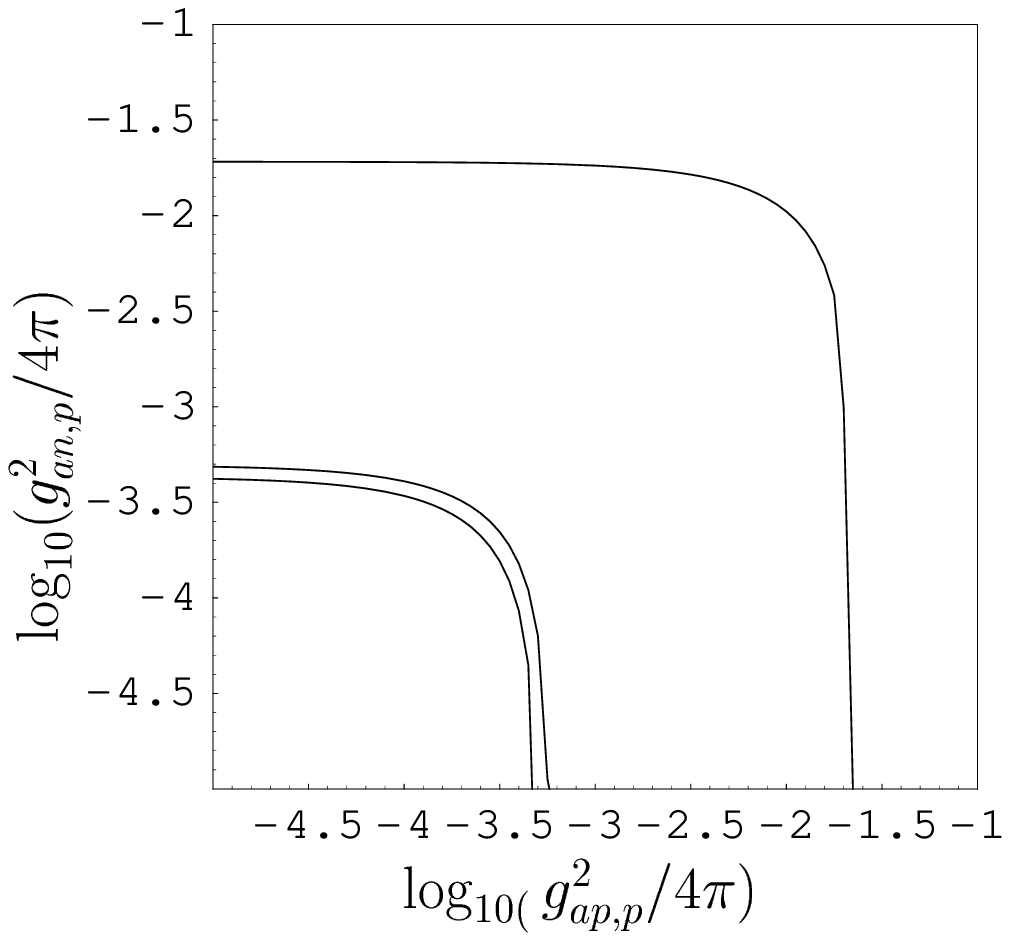}
}
\vspace*{-9cm}
\caption{Constraints on the pseudoscalar coupling constants
of an axion with a proton and a neutron following from
measurements of the thermal Casimir-Polder force are shown
as the upper, intermediate and lower lines for the axion masses
$m_a=0.2\,$eV, 0.01\,eV and $\leq 1\,$meV, respectively (see text
for further discussion). The regions of the plane above each line
are prohibited and below each line are allowed.
}
\end{figure}
\begin{figure}[b]
\vspace*{-8cm}
\centerline{\hspace*{2cm}
\includegraphics{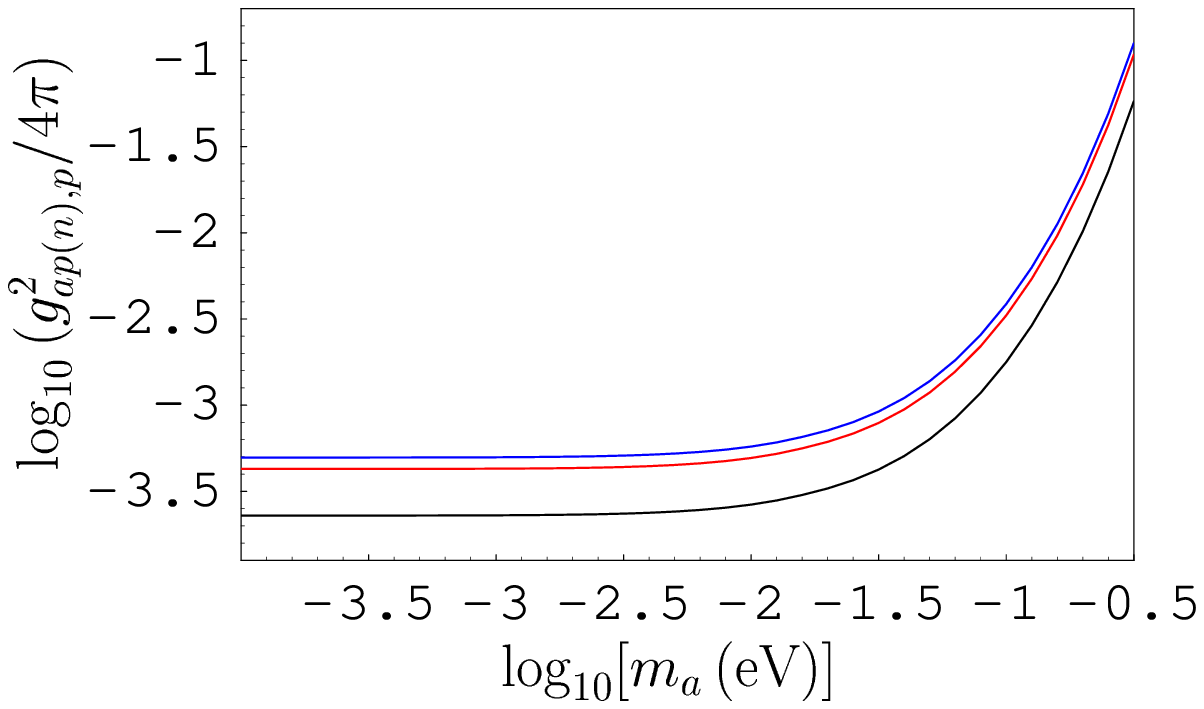}
}
\vspace*{-9cm}
\caption{(Color online) Constraints on the pseudoscalar coupling constants
of an axion with a proton or a neutron following from
measurements of the thermal Casimir-Polder force are shown
as functions of the axion mass.
The lower, intermediate and upper lines correspond to the
conditions $g_{ap,p}=g_{an,p}$, $g_{ap,p}\gg g_{an,p}$,
and $g_{an,p}\gg g_{ap,p}$, respectively.
 The regions of the plane above each line
are prohibited and below each line are allowed.
}
\end{figure}
\end{document}